# Low field magneto-transport in $La_{0.7}Ca_{0.3}MnO_3$–PMMA composites synthesized by polymeric precursor route

Jitendra Kumar[a], Rajiv K. Singh[a], H. K. Singh[a*], P. K. Siwach[b], N. D. Kataria[a], Ramadhar Singh[a] and O. N. Srivastava[b]

**a** National Physical Laboratory, Dr K S Krishnan Marg, New Delhi-110012, India
**b** Department of Physics, Banaras Hindu University, Varanasi-221005, India

## Abstract

A detailed investigation of the effect of PMMA on the structure, microstructure and magneto-transport properties of manganite $La_{0.7}Ca_{0.3}MnO_3$ (LCMO) is presented. LCMO–PMMA nanostructured composites have been synthesized by a unique polymeric sol-gel route, which leads to improved solubility of PMMA in the LCMO matrix. The LCMO phase is grown in the presence of varying PMMA concentration at ~500 $^0$C. This route yields single phase material and the grain size is observed to decrease slightly with increasing PMMA concentration. On increasing the PMMA concentration, TC undergoes a small decrease, resistivity is observed to increase by two orders of magnitude, with a concomitant large decrease in $T_{IM}$, e.g., from 218 K for virgin LCMO to 108 K for 50 wt % PMMA admixed LCMO. Low field magneto-resistance measured in the temperature range 77-300 K shows considerable enhancement as a function of the PMMA concentration. These phenomena are explained by taking into account the increased intergranular disorder as a consequence of PMMA admixture.

*Corresponding Author E-mail: hks65@mail.nplindia.ernet.in

**Introduction**

Mixed-valence perovskite manganites exhibit host of interesting physical phenomena such as colossal magneto-resistance (CMR), charge ordering (CO), orbital ordering (OO), phase separation (PS), etc. and have been the subject of intense theoretical and experimental investigations during the recent years [1, 2]. The double exchange (DE) mechanism [3] explains the intrinsic CMR effect observed around the paramagnetic (PM) to ferromagnetic (FM) transition temperature (TC). The extrinsic magneto-resistance (MR) effect observed in general in polycrystalline materials is caused either due to spin polarized tunneling (SPT) [4, 5] or spin dependent scattering (SDS) [6] as the conduction electrons traverse the grain boundaries (GBs). The dilution of GBs in manganites with insulating materials, such as a polymer, adjusts the barrier layer influencing the tunneling/scattering process which takes place across the interface/GBs and also influences the degree of magnetic disorder present therein. Enhancement of MR at low temperature and low magnetic field due to the alignment of neighboring FM grains is caused by the extraneous effects acting as pinning centers in the demagnetization by domain wall displacement. That is why a spin misorientation of the magnetically virgin state of the system is crucial to achieve and enhanced MR at low fields which is more useful for device application [7,8]. Attempts have been made to enhance low temperature-low field MR or room temperature MR by making CMR-polymer composites during the last few years [9-14]. Most commonly used polymer for making CMR-polymer composite is polyparaphenylene (PPP) [10,13]. Recently we have investigated the effect of polymethyl methaacrylate (PMMA) on a highly disordered manganite $La_{0.7}Ba_{0.1}Sr_{0.2}Mn_3$ [14]. We have observed that the solubility of the PMMA in the manganite matrix is rather low (10 wt %) and PM-FM as well as the insulator to metal (IM) transitions are destroyed rapidly accompanied by a huge increase in the resistivity. In the present investigation we have chosen 30% Ca doped $LaMnO_3$, that is, $La_{0.7}Ca_{0.3}MnO_3$ (LCMO) and synthesized the nano-sized grains of $(LCMO)_{1-x}(PMMA)_x$ (x = 0.0, 10, 20, 35 and 50 wt %) composites by a sol-gel/polymeric precursor route. The polymeric precursor/sol gel technique has efficiently been used to synthesize high quality nano-crystalline manganites [15-17]. The uniqueness of this method is that the PMMA is mixed with the LCMO precursor before the grain growth, that is, the LCMO nanocrystals are grown in the presence of PMMA and the maximum temperature (ts) used is ~ 500 $^0C$. The effect of PMMA on magneto-transport properties of $La_{0.7}Ca_{0.3}MnO_3$ is reported in the present paper.

**Experimental**

LCMO-PMMA composites were prepared by polymeric precursor route [15]. In the first step of synthesis, the relevant metal nitrates viz., La(NO3)36H2O, Ca(NO3)24H2O and Mn(NO3)24H2O (purity > 99 %) were dissolved in deionized water in the desired cationic ratio La : Ca : Mn = 0.7 : 0.3 : 1. After the proper homogenization of the aqueous solution, an equal amount of ethylene glycol was added. The resulting solution was thoroughly homogenized and then continuously stirrered at ~1500 C. The heat treatment was continued till the solvents dried and a resin was formed. This residual resin was further decomposed by heat treatment at ~2500 C. The decomposition of the resin yields slightly blackish foam like substance which acts as a precursor for the growth of polycrystalline LCMO and LCMO-PMMA composites. This foam is then crushed and pressed in form of pellets which is heat treated in air at ~5000 C for 8 hrs to get the nano-crystalline LCMO. The LCMO-PMMA composites were prepared by the following process. PMMA was weighed in the desired amounts (10, 20, 35 and 50 wt %) and dissolved in chloform. Then the LCMO precursor was added to the chloroform-PMMA solution and mixed thoroughly till the whole of the chloroform evaporated. The resulting powder mixture was dried at ~150 0C and then pressed in the form of rectangular pellets of 10x5x1 mm3 dimension. These pellets were then heated in air at ~5000 C for 8 hrs to get the LCMO-PMMA nano-crystalline composites. LCMO phase in the composite is grown in the presence of PMMA, the homogeneity of the composites is expected to be better, because of the amorphous nature of the LCMO precursor foam obtained at ~250 °C. The LCMO samples having polymer concentration x = 0.0, 10, 20, 35 and 50 wt % will hereafter be referred to as LCMP0, LCMP10, LCMP20, LCMP35 and LCMP50 respectively. These LCMO-PMMA composite samples were characterized by powder Xray diffraction (XRD), scanning electron microscopy (SEM), resistivity ($\rho$); low field magneto-resistance (LFMR) and ac susceptibility ($\chi$) measurements in the temperature range 77-300 K.

**Results and Discussion**

Figure 1 shows the X-ray diffraction pattern of LCMP0, LCMP10, LCMP20, LCMP35 and LCMP50, respectively. This figure reveals that all the samples are polycrystalline and have orthorhombic unit cell. The lattice parameters of the virgin LCMO are; a = 5.436 Å, b = 5.482 Å and c = 7.675 Å and a small increase is observed in PMMA admixed samples which, however, saturates at higher PMMA concentrations. The reflected intensity is observed to decrease with increasing PMMA concentration with a concomitant increase in the full width at half maximum (FWHM) of the corresponding reflections. Figure 2 depicts the close up of the (110) reflection of LCMP0, LCMP20 and LCMP50 samples. The increase in FWHM suggests a slight decrease in the LCMO grain size in the composite or may be suggestive of the presence of PMMA in the background. As expected, this suggests a decrease in the degree of crystalline nature with increasing PMMA concentration. As evidenced by shifting of the (110) reflection of LCMP20 and LCMP50 towards lower 2θ, the lattice parameters of the orthorhombic unit cell show small increase with PMMA concentration beyond 10 wt %. The orthorhombicity of the system usually defined by the expression Or(%) = [(b-a)/(b+a)]*100, is observed to be 0.421 % for LCMP0 and is almost not affected in the composites. The average grain diameter/grain size (p) was evaluated indirectly from the XRD data by applying the Scherrer formula p = 0.9λ/βcosθ using the FWHM (full width at half maximum) of the (110) reflection. The same was also determined directly by employing scanning electron microscopy (SEM) to image the surface microstructure of the LCMO-PMMA composite samples. Figure 3 (a-d) shows the scanning electron micrographs of LCMP0, LCMP20, LCMP35 and LCMP50. The micrographs reveal a small variation in the grain size with increase in the PMMA concentration. The average grain size is found to be ~ 40 nm in the virgin LCMO sample and size does show a small decrease as the PMMA concentration is increased. This feature is clearly observed in the SEM micrograph of LCMP50 (Figure 3 (d)).

The PM-FM phase transition studied by temperature variation of ac susceptibility (χ-T) measurement in the temperature range 77-300 K also reveals some interesting aspects of LCMO-PMMA composites. The χ-T data of all the samples is shown in Figure

4. The TC of the samples was determined from the peak in the d$\chi$/dT curve (results not shown) and is representative of the mid point of the transition. The virgin LCMO sample (LCMP0) undergoes a PM-FM transition at $T_C$ ~ 261 K while all the composite samples have successively lower $T_C$ values. The $T_C$ of LCMP10, LCMP20, LCMP35 and LCMP50 are ~ 256, 255, 244 and 236 K, respectively. In addition to the successive lowering of the $T_C$ with increasing PMMA concentration, the transition width is also observed to increase appreciably. This suggests that despite the non-magnetic nature of PMMA, its admixture to LCMO results in partial decoherence of the FM domains. Since PMMA can not be incorporated into the LCMO matrix it diffuses and segregates into the GBs/interfacial regions. These GBs/interfacial regions are known to have both magnetic and structural disorder. The most common disorder being Mn spins blocked at GBs, increased anisotropy in the interfacial regions and misalignment of the magnetic moments of the neighboring FM domains. The addition of PMMA, despite its nonmagnetic character, is expected to further increase the density of these disorders. The increased density of these disorders means the widening of the GB regions and thus the system can be considered a phase separated one, having a mixture of FM (metallic) phase and the non-magnetic (insulating) phase. As the PMMA concentration is increased, the overall density of the non-magnetic/insulating region in the bulk increases, resulting in the partial decoherence of the long range FM order. Since even at 50 wt % PMMA the FM transition is still as high as 236 K, it is conjectured that the fundamental/intrinsic mechanism of the FM transition, viz., the double exchange, is not modified appreciably due to the PMMA admixture. The increased density of the disorders described above is expected to lead to large enhancement in the dc resistivity of the LCMO-PMMA composites with increasing polymer concentration.

The dc resistivity ($\rho$) was measured in the temperature range 77-300 K by four probe technique and the data are plotted in Figure 5. At room temperature (~298 K), resistivity values are ~3.75, 6.05, 34.94, 49.35 and 90.13 $\Omega$-cm, respectively for LCMP0, LCMP10, LCMP20, LCMP35 and LCMP50. Thus even at room temperature large enhancement in the resistivity is observed as a consequence of PMMA admixture. This resistivity enhancement is caused mainly by the scattering of the charge carriers by the magnetic disorder in the GBs and as the PMMA concentration increases more and more scatterers are produced resulting in the further increase in the resistivity. The lowering of the temperature towards the $T_C$ causes an increase in the resistivity and at a certain temperature which is lower than $T_C$ an insulator ($d\rho/dT < 0$) to metal (($d\rho/dT > 0$) like transition is observed in all the samples. The measured insulator-metal transition temperature ($T_{IM}$) are ~218, 213, 173, 153 and 108 K, respectively for LCMP0, LCMP10, LCMP20, LCMP35 and LCMP50 and the corresponding $\rho$ values are ~15.92, 26.56, 227.97, 432.25 and 3832.51 $\Omega$-cm, respectively. The variation of $T_C$ and $T_{IM}$ with the PMMA concentration is plotted in Figure 6 where as the resistivity dependence on PMMA concentration at different temperatures is depicted in Figure 7. The large difference in the $T_C$ and $T_{IM}$ for the virgin LCMO sample (~ 43 K) is due to the existence of the disorder and is in fact a common feature of the polycrystalline manganites having small grain size [1]. The strong suppression of the $T_{IM}$ as compared to $T_C$ is caused by the PMMA induced disorders and also by the increase in the non-magnetic phase fraction. This also causes the increase in the carrier scattering leading to a corresponding enhancement in the resistivity. In fact, the admixture of non-magnetic and insulating PMMA separates the FM metallic clusters and as the polymer concentration increases the spatial separation of these grains/clusters further increases. This leads to the lowering of the metallic transition temperature and concomitant enhancement in the resistivity. When a magnetic field is applied, the FM clusters grow in size and the interfacial Mn spin disorder is suppressed resulting in the improved connectivity and consequently a decrease in the resistivity.

The temperature dependence of MR measured in the range 77-300 K at 2 kOe is shown in Figure 8. The nearly monotonic temperature dependence of MR below $T_C$ suggests that in LCMP0 and LCMP10 samples, there is no contribution of the intrinsic component of MR which arises due to the double exchange mechanism around $T_C$. However, around $T_C$ the jump in the MR-T curve of the LCMP0 as well in LCMP10 shows small contribution from the intrinsic component to the total low field MR. In the PMMA admixed LCMO samples the jump in the MR around $T_C$ shifts towards lower temperatures, (for example - LCMP10 and LCMP20) and finally it disappears when the PMMA concentration is increased above 20 wt %. This is due to the fact that as the PMMA concentration is increased the intrinsic contribution to the low field MR is suppressed and as a consequence of this the MR of all composite samples is lower than that of the virgin LCMO sample till a certain lower crossover temperature is reached. The MR crossover temperatures are ~150, 144 and 120 K for LCMP20, LCMP35 and LCMP50, respectively. Below these temperatures, the LCMO-PMMA composite samples exhibit larger MR than the virgin LCMO sample. At 77 K, the MR values are measured to be ~ 13.1, 13.7, 14.8, 15.5 and 16.4 % for LCMP0, LCMP10, LCMP20, LCMP35 and LCMP50, respectively. Thus, increasing the PMMA concentration leads to the enhancement in low field MR at lower temperatures while the MR in the higher temperature regime is suppressed. The disappearance of the MR in the higher temperature regime can be explained by taking into account the successive dilution of the FM magnetic phase and the double exchange (DE) mechanism around the respective FM transition temperatures as a consequence of increasing PMMA concentration.

The magnetic field dependence of the low field MR at 77 K for all the samples are given in Figure 9. The low field MR of virgin LCMO as well as LCMO-PMMA composite samples is observed to increase with increasing magnetic field. The MR-H curve shows two different slopes, the one below H~1 kOe is steeper while the other above H~1.5 kOe is rather weak. As the PMMA concentration is increased MR increases and the slope of the MR-H curve in both the temperature regimes becomes steeper. In the virgin LCMO sample, the low field MR at H = 0.8 kOe and 3.6 kOe is measured to be ~ 9 % and ~15 %, respectively and the same in the case of LCMP50 increases to ~11 % and ~20 %, respectively. Thus as the polymer concentration is increased to 50 wt % PMMA, the relative enhancement in MR shows an increase of ~ 20 % in the lower field regime and ~35 % in the higher field regime. It may also be noted that as the PMMA concentration increases, the saturation of MR with magnetic field disappears and hence the slope of the MR-H curve in the higher magnetic field regime (H>1.5 kOe) increases.

The variation in MR as a function of PMMA concentration can be explained by taking into account the increased disorder in the inter-granular regions. The grain boundaries in polycrystalline manganites mimic the role of the thin insulating layer sandwiched between two FM grains and therefore the electron hopping across the grain boundaries depends essentially on these GBs themselves and the spin states in the neighboring grains. It is known that in polycrystalline manganites the interfacial/inter-granular regions are magnetically as well as structurally disordered and hence have disordered Mn spins. In addition, certain degree of spin canting may also be present. When a non-magnetic impurity such as PMMA is added these disorders are further increased. As the spin disorder increases, there is a concomitant increase in the carrier scattering and consequently the resistivity of the samples increases. Such variation has indeed been observed in the present investigation and the resistivity has been observed to increase with increasing PMMA concentration (increased magnetic disorder is proportional to PMMA concentration). The Mn spin disorder is suppressed under the influence of an external magnetic field causing a decrease in the resistivity and hence increase in MR. The increase in the slope of the MR-H curve in the higher field regime as a function of PMMA concentration also suggests that the spin disorder created (due to increasing PMMA concentration) has not been fully suppressed by a magnetic field H = 3.6 kOe. Consequently no MR saturation (as observed in virgin LCMO (LCMP0) and 10 wt % PMMA admixed LCMO (LCMP10)) is seen in other composite samples having higher PMMA concentration. In fact as Mn spin disorder density increases due to PMMA admixture, the carrier scattering increases leading to further enhancement in the resistivity and the magnetic interaction energy also increases and hence a higher magnetic field is needed to suppress them.

**Conclusions**

In the present investigation we have studied the (LCMO)1-x(PMMA)x, (x = 0.0, 10, 20, 35 and 50 wt %) composites synthesized by a unique polymeric sol-gel route that facilitates the LCMO grain growth in presence of the polymer. All the samples are single phase and nano-structured, and a small decrease in the average grain diameter is evidenced by XRD and SEM results. The PM-FM transition is reduced moderately. The resistivity increases nearly by two orders of magnitude and the insulator-metal transition; $T_{IM}$ is severely suppressed when the PMMA concentration is increased to 50 wt %. The intrinsic component of MR is reduced while the LFMR is observed to increase gradually with the polymer concentration and at 77 K it shows a relative enhancement of nearly 35 % at a magnetic field ~3.6 kOe. The observed variation in the magneto-transport parameters have been explained in terms of the magnetic disorder created due to PMMA admixture to LCMO.

**Acknowledgements**

The authors are thankful to Professor Vikram Kumar, Director, National Physical Laboratory, New Delhi for his keen interest and permission to publish this work. Two of us (J. K. and P. K. S.) are thankful to Council of Scientific and Industrial Research, Government of India, New Delhi for the award of Senior Research Fellowship.

**Figure Captions**

Figure 1  Powder XRD pattern of $(LCMO)_{1-x}(PMMA)_x$ (x=0.0, 10, 20, 35, 50) composites.

Figure 2  Close up of the (110) reflection of virgin LCMO (LCMP0), 20 wt 5 % (LCMP20) and 50 wt % (LCMP50) PMMA admixed LCMO.

Figure 3  SEM micrographs showing the surface microstructure of virgin (a) LCMP0, (b) LCMP20, (c) LCMP35 and (d) LCMP50.

Figure 4  Temperature dependent ac susceptibility of $(LCMO)_{1-x}(PMMA)_x$ (x=0.0, 10, 20, 35, 50) composites.

Figure 5  Temperature dependence of resistivity of $(LCMO)_{1-x}(PMMA)_x$ (x=0.0, 10, 20, 35, 50) composites at zero magnetic field.

Figure 6  Variation of the PM-FM transition temperature ($T_C$) and the insulator-metal transition temperature ($T_{IM}$) of $(LCMO)_{1-x}(PMMA)_x$ (x=0.0, 10, 20, 35, 50) composites with PMMA concentration.

Figure 7  Variation of the resistivity of $(LCMO)_{1-x}(PMMA)_x$ (x=0.0, 10, 20, 35, 50) composites with PMMA concentration at different temperatures.

Figure 8  Temperature dependence of LFMR of $(LCMO)_{1-x}(PMMA)_x$ (x=0.0, 10, 20, 35, 50) composites measured at dc magnetic field H = 2 kOe.

Figure 9  Magnetic field dependence of LFMR of $(LCMO)_{1-x}(PMMA)_x$ (x=0.0, 10, 20, 35, 50) composites measured at 77 K.

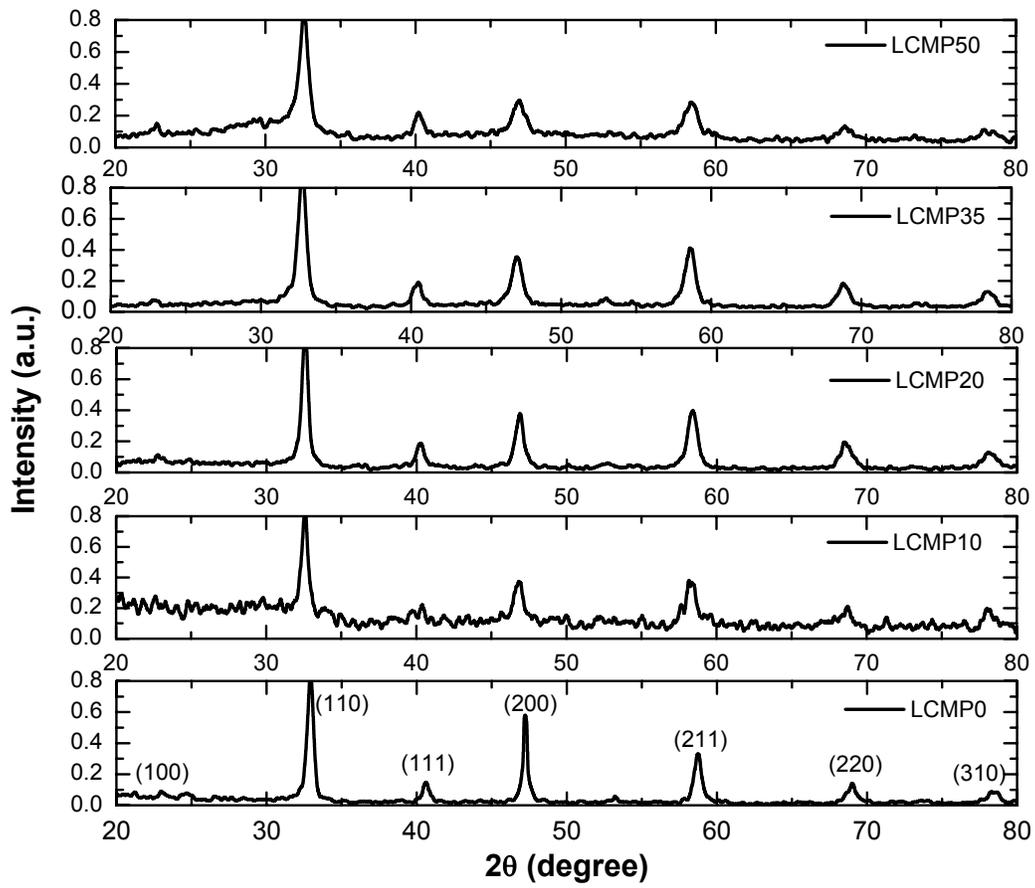

**Figure 1.**

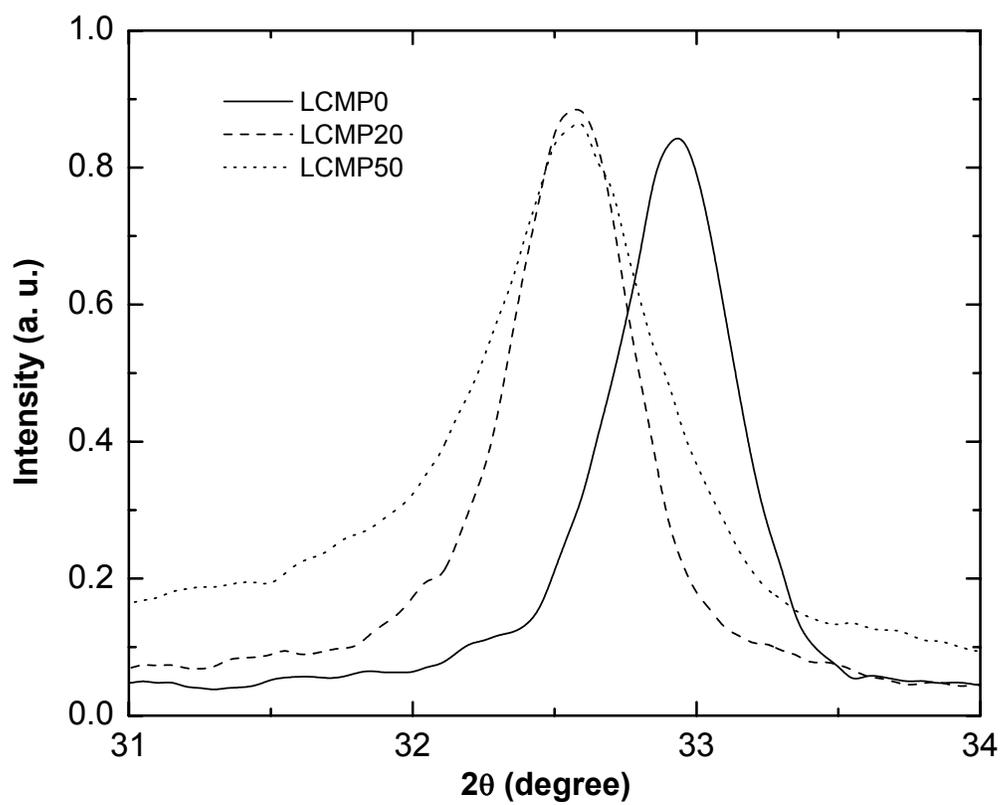

**Figure 2.**

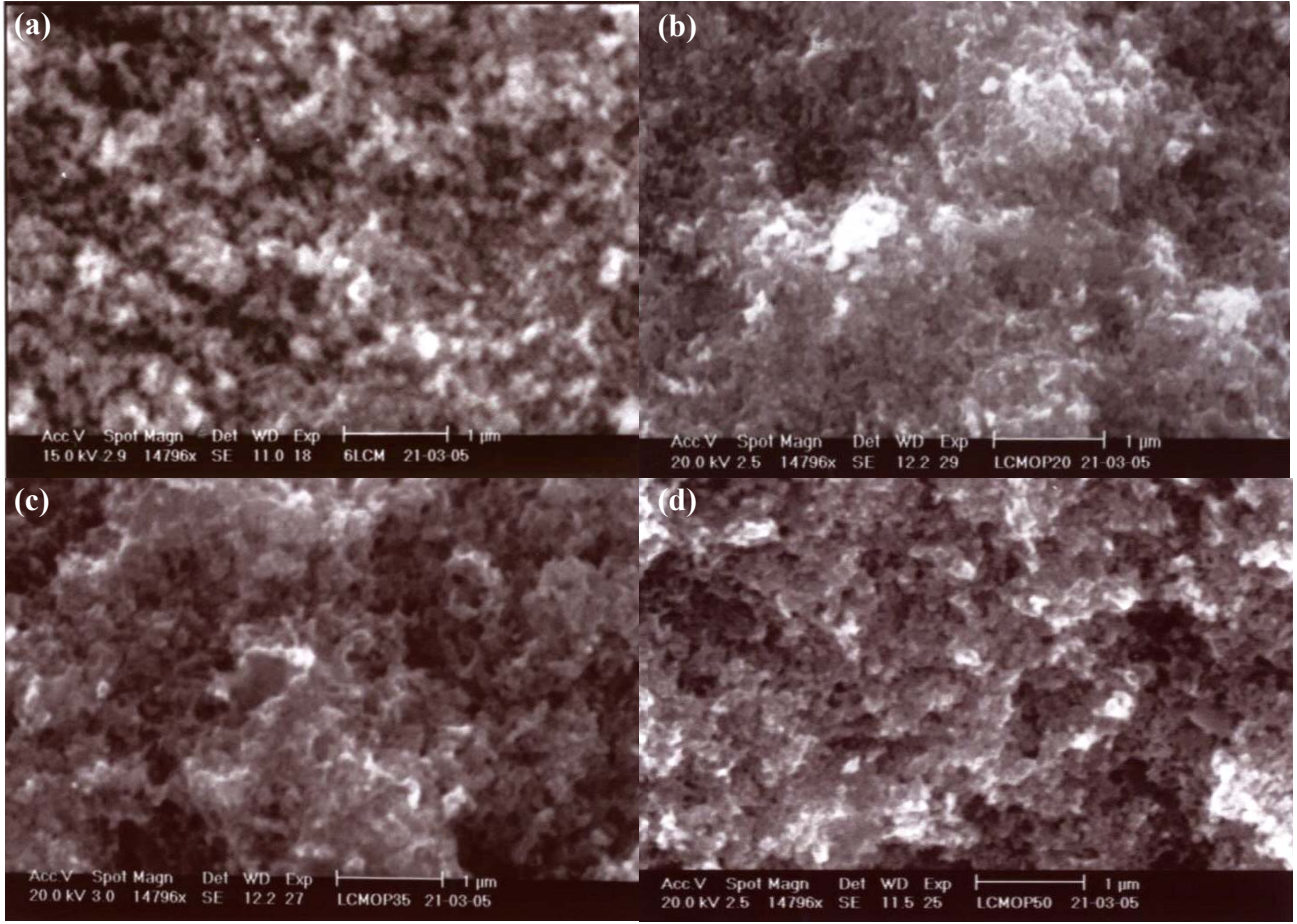

**Figure 3**

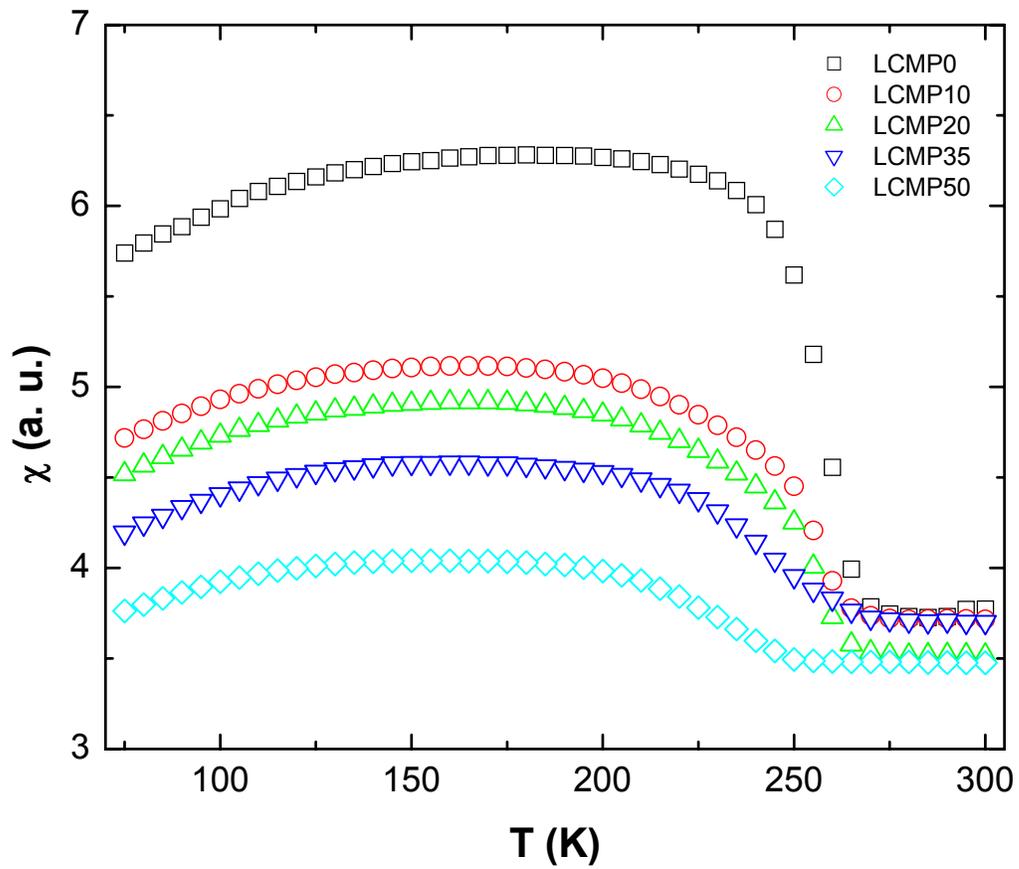

**Figure 4**.

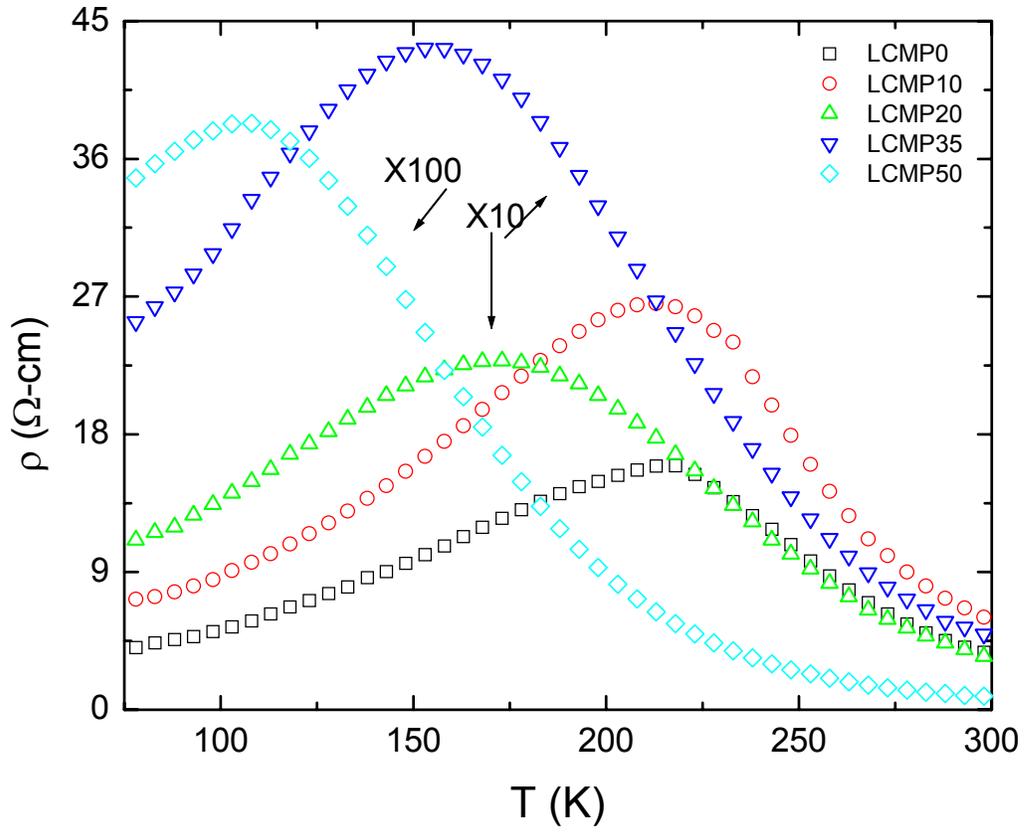

**Figure 5.**

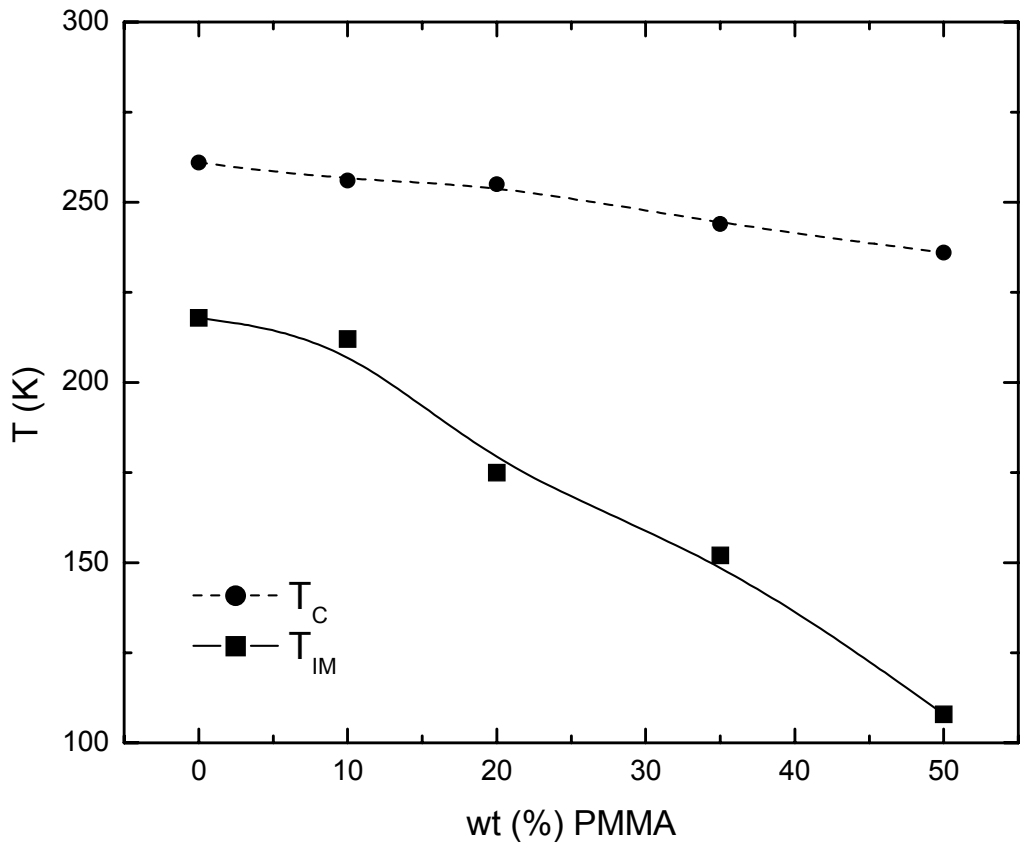

**Figure 6.**

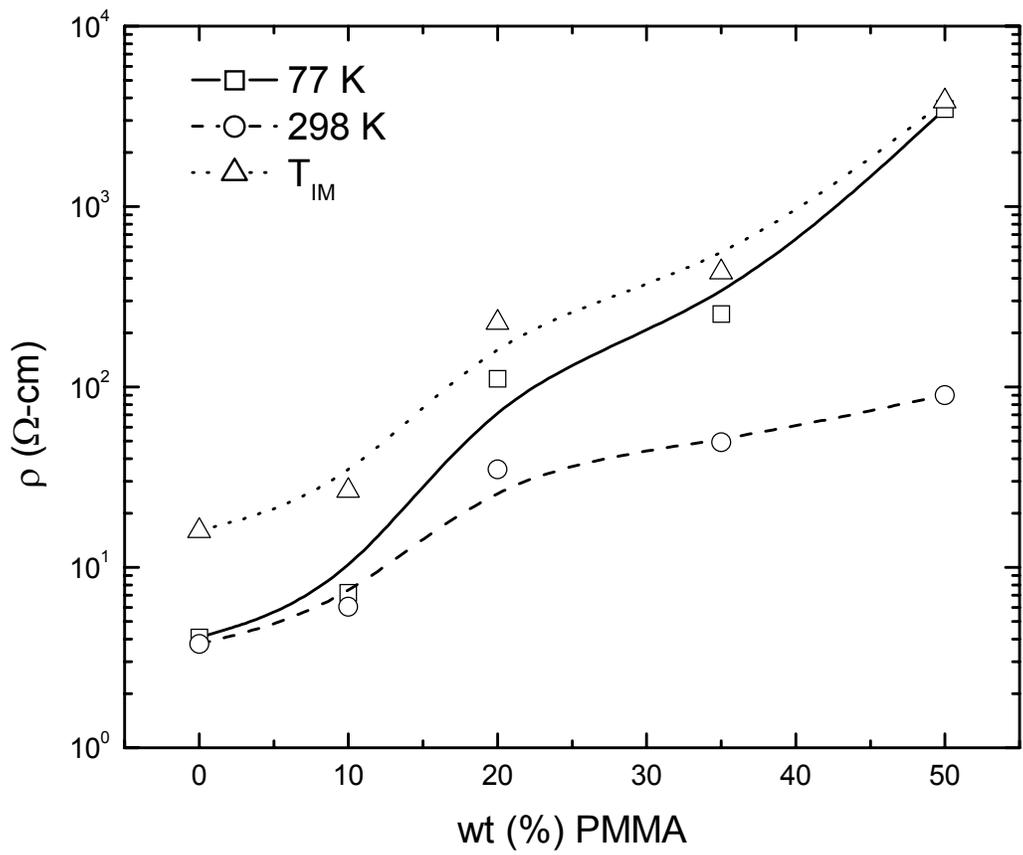

**Figure 7.**

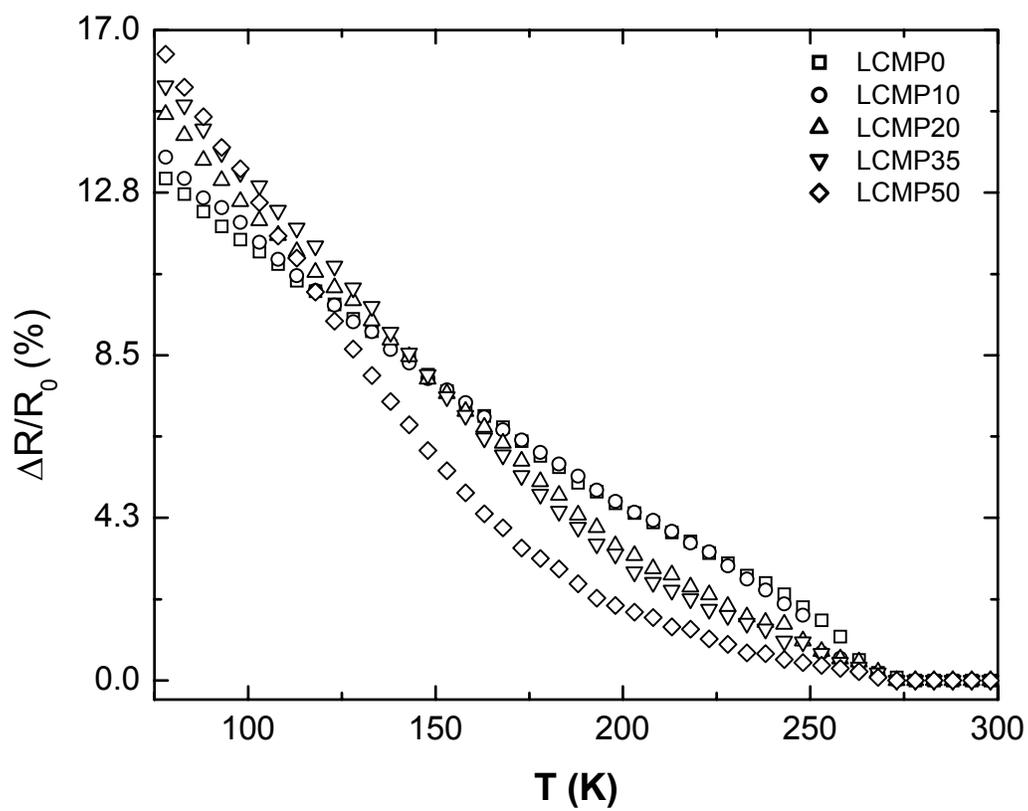

**Figure 8.**

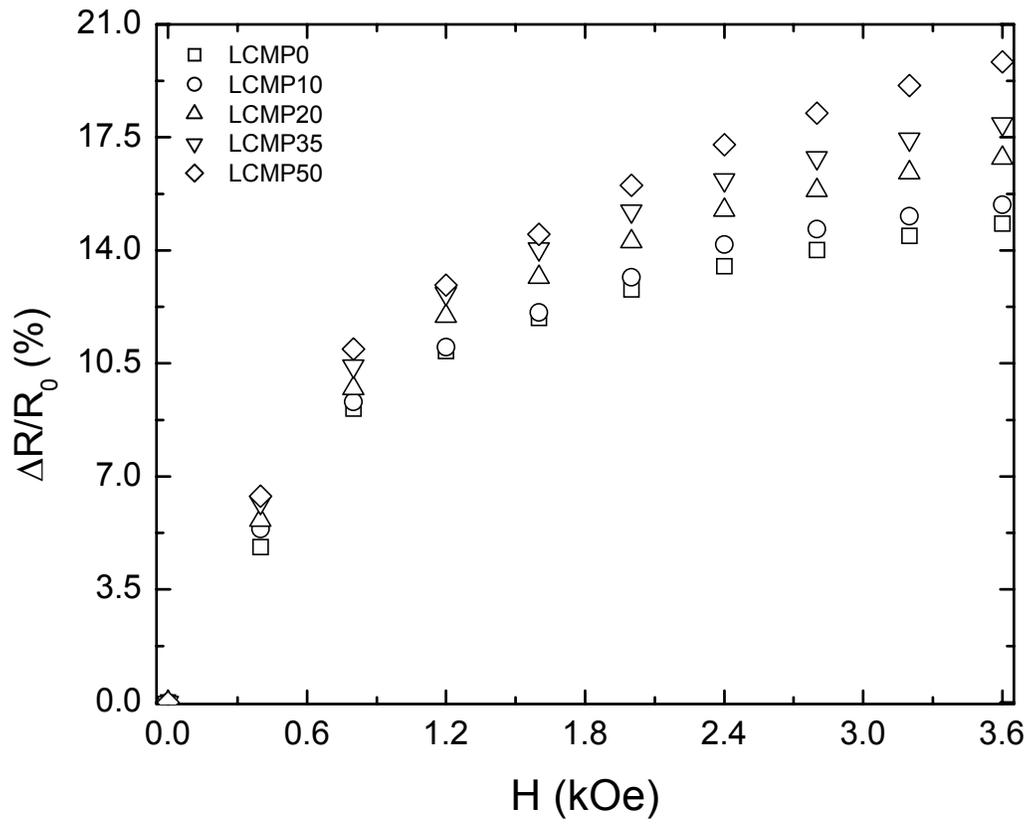

**Figure 9.**